\documentclass[10pt,preprintnumbers,twocolumn,amsmath,amssymb,nofootinbib,superscriptaddress]{revtex4-2}

\usepackage[latin1]{inputenc}
\usepackage{slashed}
\usepackage{amsmath}
\usepackage{textcomp}
\usepackage{amssymb}
\usepackage{amsfonts}
\usepackage{indentfirst}
\usepackage{color}
\usepackage{xcolor}
\usepackage{hyperref}
\usepackage{subcaption}
\usepackage[shortlabels]{enumitem}
\usepackage[normalem]{ulem}
\usepackage{comment}
\usepackage{float}
\usepackage{ragged2e}

\newcommand{\be}{\begin{equation}}
\newcommand{\ee}{\end{equation}}
\newcommand{\bea}{\begin{eqnarray}}
\newcommand{\eea}{\end{eqnarray}}

\usepackage{graphicx,graphics}
\graphicspath{{figures/}}

\begin{document}

\hfill{KCL-PH-TH/2023-42}

\title{Tunnelling-induced cosmic bounce in the presence of  anisotropies}

\author{Jean Alexandre} 
\affiliation{Theoretical Particle Physics and Cosmology, King's College London, WC2R 2LS, United Kingdom}
\author{Katy Clough}
\affiliation{Geometry, Analysis and Gravitation, School of Mathematical Sciences, Queen Mary University of London, Mile End Road, London E1 4NS, United Kingdom}
\author{Silvia Pla}  

\affiliation{Theoretical Particle Physics and Cosmology, King's College London, WC2R 2LS, UK}

\begin{abstract}
If we imagine rewinding the universe to early times, the scale factor shrinks and the existence of a finite spatial volume may play a role in quantum tunnelling effects in a closed universe. It has recently been shown that such finite volume effects dynamically generate an effective equation of state that could support a cosmological bounce. In this work we extend the analysis to the case in which a (homogeneous) anisotropy is present, and identify a criteria for a successful bounce in terms of the size of the closed universe and the properties of the quantum field. 
\end{abstract}

\maketitle

\section{Introduction}

Our universe is expanding, and on the largest scales it appears homogeneous and isotropic with a small spatial curvature \cite{Planck:2018jri}. In the standard paradigm, in which our universe emerged from a cosmological singularity, inflation provides a dynamical mechanism to achieve the current state for a generic initial condition \cite{Guth:1980zm,Linde:1981mu,Albrecht:1982wi}. However, the success of inflation is not entirely independent of the initial conditions - some models require a certain level of homogeneity in order to proceed (see \cite{Brandenberger:2016uzh} for a review), and all inflationary potentials will fail to create an exponential expansion for an initially collapsing state in the absence of a violation of the Null Energy Condition (NEC) \cite{Hawking:1970zqf,Borde:1993xh,Rubakov:2014jja,Kontou:2020bta}. 
An alternative paradigm, that of ekpyrosis \cite{Khoury:2001wf,Khoury:2001bz,Steinhardt:2001st}, commonly uses a mechanism of slow contraction to provide the smoothing of inhomogeneities in the case of a non singular cosmic bounce, (see \cite{Ijjas:2018qbo} for a review). Such models also necessitate a violation of the NEC in order to transition to expansion. Therefore mechanisms that violate the NEC in the early universe are of interest for such scenarios.

Most mechanisms for NEC violation require additional exotic components or a modification of general relativity (GR) \cite{Rubakov:2014jja}. 
Recently, a mechanism has been proposed in which the NEC is violated by finite volume effects, which necessarily occur for a scalar field with a Higgs-like potential in standard Quantum Field Theory (QFT) on an FLRW background with a closed topology \cite{Alexandre:2023pkk}. The effect arises from tunnelling between two degenerate vacua, which is allowed if the field is confined in a finite spatial volume
\footnote{Note that this effect is distinct to the well known Casimir effect \cite{Bordag:2001qi}.
Although both are based on the finite-volume assumption, tunnelling is independent of the geometry/topology of the spatial unit cell. 
For a comparison between the two effects, see \cite{Alexandre:2023iig}.}.
One nice aspect of this mechanism is that it ``turns off'' in a period of expansion, meaning that after a cosmological bounce it would quickly become suppressed - it therefore naturally favours expansion over contraction. 
We note here that alternative quantum effects, involving fermion dynamics, have been proposed to induce NEC-violation and potentially lead to a cosmological bounce \cite{Trautman:1973wy,Alexander:2008vt,Magueijo:2012ug,Tukhashvili:2023itb}.

A key question is whether the mechanism described in \cite{Alexandre:2023pkk} could also provide some kind of smoothing of inhomogeneities or anisotropies, and to what extent it must dominate over these in order for the bounce to be successful. In this work we will discuss the (homogeneous) anisotropic case, and explain why the energy of the quantum fluid must already dominate over the anisotropy before the bounce in order for it to proceed (effectively, this is just the requirement that the equation of state parameter $w<1$ to avoid chaotic mixmaster behaviour from dominating \cite{Erickson:2003zm}). As in other bounce scenarios, a scalar field that is dominated by its kinetic energy or other stiff fluid with equation of state $w \ge 1$ could play the role of a smoother in a preceding slow contraction phase \cite{Ijjas:2018qbo}. This then requires some transition between such a component and the effects that cause the bounce dominating - in our model, for example, it may be possible that the field itself could be responsible for smoothing at an earlier kinetic dominated field (for example, as it rolls down into one of the minima of the potential, at which point the tunnelling effects dominate) 
\footnote{Also, gravitational particle creation tends to rapidly suppress irregularities in the geometry, 
which can be seen with semiclassical backreaction effects \cite{Hu:2021pfh}.}.
However, the description we use here is only valid in the vicinity of the bounce, and so more work is required to quantify out-of-equilibrium effects and the impact of inhomogeneities at an earlier stage. These aspects are more difficult to treat and need to be explored further in future work, as well as considering the possible origins of the fluctuations that are observed on larger scales, and their consistency with observations from the CMB and other cosmological probes. In this work we simply assume that by the time the universe is nearing the bounce, the anisotropies are suppressed, and consider what this implies for the properties of the quantum field that must drive the process.

We will show that criteria for the success of the bounce can be stated in terms of the size of the closed universe at the point at which the net energy density is zero, and the properties of the quantum field (mainly its mass and vacuum energy). We focus on the case of anisotropy as a second component since it is the component that dominates the energy budget the quickest during a collapse, but we could equally have considered other secondary components with other equations of state. Roughly speaking, at the point of zero net energy density, the size of the (closed) universe must be comparable to the Compton wavelength of the field for its pressure to be sufficient to turn around the collapse. We will make this statement more precise in what follows, and give the phenomenological consequences for the field in assuming that this closed universe size is equal to or greater than that of the observable universe.

We note that our model arises from the quantisation of a scalar field with different classical configurations, on a classical background metric, unlike studies involving the mini-superspace approach, which provide a toy model for Quantum Gravity. In the latter, the path integral can also be dominated by  different classical configurations \cite{Halliwell:1988ik, Feldbrugge:2017kzv,DiTucci:2019xcr}, but for the metric rather than an additional scalar. Anisotropies have also been discussed in the mini-superspace context \cite{Rajeev:2021yyl}. Furthermore, bouncing cosmological models have been extensively reviewed in \cite{Barca:2021qdn} in the context of Loop Quantum Cosmology and Polymer Quantum Mechanics  (see also \cite{Barca:2021epy} for a comparison of different models in Bianchi I spacetime).


The article is organised as follows: In Sec. \ref{Sec:bg_tunnelling}, we summarise the background of tunnelling in a finite volume, in Sec. \ref{Sec:bg_aniso_cosmo} we set out the standard description of a homogeneous anisotropic cosmology, in Sec. \ref{Sec:aniso_qft} we extend the QFT description to the anisotropic case and in Sec. \ref{Sec:analytic} we describe the conditions for success in terms of the model parameters and discuss phenomenology. We briefly provide some numerical illustrations in Sec. \ref{Sec:numeric} and conclude with a brief discussion in Sec. \ref{Sec:discuss}.

\section{Background I: Tunnelling in an FLRW background with finite volume}
\label{Sec:bg_tunnelling}

\begin{figure}[t] 
 \includegraphics[width=8.6cm]{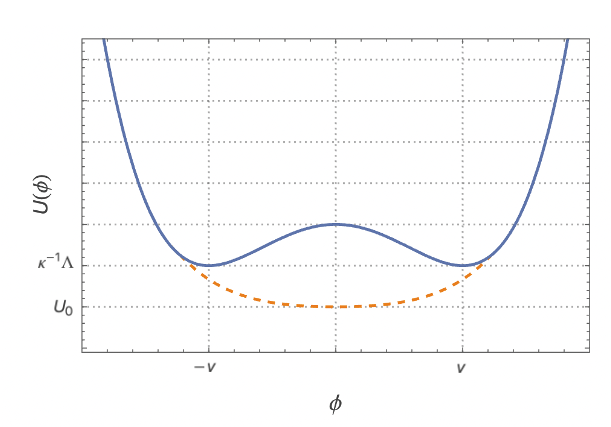}  
        \caption{\justifying{\small{Schematic representation of the bare potential $U(\phi)$ (blue line) and the effective potential $U_{eff}(\phi)$ 
        obtained from tunnelling between the vacua (orange dashed line). $U_0$ is given by eq.(\ref{U0}), from which we can see that the effective potential curve reaches a lower value for a smaller volume. In this way the finite volume effects act as a negative contribution to the energy density, and can violate the NEC during a period of contraction. }}}
        \label{potential}
\end{figure}

Spontaneous Symmetry Breaking, where the scalar field is trapped above one vacuum, is only strictly valid in an infinite volume, where tunnelling to another degenerate vacuum is completely suppressed. In a finite volume, tunnelling between two degenerate bare vacua $\phi=\pm v$ is possible, leading to an effective potential with a lower overall minimum (see Fig. \ref{potential}).

Using a semi-classical approximation for the partition function, 
which is dominated by a dilute gas of instantons and anti-instantons, it was shown in 
\cite{Alexandre:2022qxc,Alexandre:2023iig} that the resulting effective theory is such that:
\begin{enumerate}
    \item the true vacuum is symmetric and occurs at $\phi=0$, consistently with convexity of the effective potential when several saddle points are taken into account
    \cite{Symanzik:1969ek,Coleman:1974jh, Iliopoulos:1974ur, Fujimoto:1982tc, Haymaker:1983xk,Bender:1983nc, Hindmarsh:1985nc, Alexandre:2012ht, Plascencia:2015pga, Millington:2019nkw,Alexandre:2022sho};
    \item the corresponding effective action is not extensive (it is not proportional to the volume), but has a non-trivial volume dependence;
    \item the resulting ground state fluid violates the NEC.  
\end{enumerate}
The above features arise from non-perturbative properties of the partition function, since a convex effective potential cannot be obtained from a bare double-well potential with perturbative quantum corrections only.

The (anti-)instantons considered in  \cite{Alexandre:2022qxc,Alexandre:2023iig} depend only on the Euclidean time, and the tunnelling process is similar to the one described in Quantum Mechanics, where tunnelling of a particle in a double-well potential leads to a true ground state energy which is lower than the ground state energy in both individual degenerate wells \cite{Kleinert:2004ev}
\footnote{Because the vacua are degenerate, the $O(4)$-symmetric Coleman bounce \cite{Coleman1,Coleman2} does not play a role here, since it would require a bubble with infinite radius.
The tunnelling mechanism described in \cite{Alexandre:2022qxc,Alexandre:2023iig} is therefore not related to a first order quantum phase transition, but rather to a second order phase transition and happens uniformly in space.}.

Following the same approximation for the partition function, but in a flat and isotropic Friedmann-Lemaitre-Robertson-Walker universe (FLRW),
it was shown in \cite{Alexandre:2023pkk} that the above mechanism 
dynamically generates a cosmic bounce, which is followed by an asymptotic de Sitter phase where tunnelling is suppressed exponentially. 
In this context, ``finite volume'' is provided by a unit cell in the form of a 3-torus of volume $V_{cell}$.
Some recent works in cosmology have considered the evidence for a closed universe,  see for example \cite{DiValentino:2019qzk,2021PhRvD.103d1301H}.
However, since we do not see any periodicity in our observable universe, any closed volume must be larger than its current size, so any finite volume effects will now be negligible. The physical size of the closed universe can of course be much smaller far in the past, when our comoving volume was smaller, and thus finite volume effects could have played a role in the early universe.

In the isotropic case of \cite{Alexandre:2023pkk}, tunnelling between the minima of the double-well potential 
\be
U(\phi)=\kappa^{-1}\Lambda_b+\frac{\lambda_b}{24}(\phi^2-v_b^2)^2~,
\ee
leads to the convex effective potential 
\be\label{convexU}
U_{eff}(\phi)=U_0+\frac{1}{2}M^2\phi^2+{\cal O}(\phi^4)
\ee
where
\bea\label{U0}
U_0&=&\kappa^{-1}\Lambda\left(1-r\frac{e^{-\alpha_{iso}^3}}{\alpha_{iso}^{3/2}}\right)\\
M^2&=&\frac{\lambda v^2}{3}\left(1-\frac{27\lambda }{32\pi^2}\right)>0~.\nonumber
\eea
In the above expressions, $\lambda_b$, $\Lambda_b$ and $v_b$ have to be understood as the bare parameters, while $\Lambda$, $\lambda$ and $v$ are the renormalised parameters
\footnote{For renormalisation purposes, it is necessary to introduce the bare vacuum energy $\Lambda_b$. The renormalised vacumm energy $\Lambda$ arises from the scalar field self-interactions; it is not put by hand but it is generated by quantum fluctuations. The latter are dominated by ultraviolet effects  and not infrared effects, such that $\Lambda$ can be assumed to be independent of $V_{cell}$.}. 
Also, $\kappa=8\pi G$ and we have defined
\be\label{rdef}
r=\frac{\lambda \kappa v^4}{3\sqrt{3\pi}~\Lambda}~~~~,~~~~ 
\alpha_{iso}^3\equiv a^3\Sigma~,
\ee 
where $\Sigma$ is the action of one instanton relating the bare vacua and $a$ is the FLRW scale factor (we consider $\hbar=c=1)$. 
The potential is illustrated in Fig. \ref{potential}.

Both $r$ and $\Sigma$ depend on the field parameters, but $\Sigma$ is also proportional to the volume $V_{cell}$ of the fundamental spatial cell, which therefore needs to be finite for the tunnelling probability $\propto\exp(-\alpha_{iso}^3)$ to be finite - that is, one requires a closed universe. 

The present study extends the work of \cite{Alexandre:2023pkk} to the anisotropic case, and considers the impact of other components being present. In this work, an adiabatic approximation is assumed, where the tunnelling rate is large compared to the expansion rate, 
which allows the use of equilibrium QFT. 
This approximation is very good in the vicinity of the bounce, which is the regime on which we focus. 

 Finally, we comment here on the stability of our solution against small fluctuations of the spatial curvature $^{(3)}R\ll v^2$. For $\xi=0$, the bare potential is not modified and no change would occur in our results. For $\xi \neq0$ the presence of a small non-vanishing curvature would slightly shift the position of the true vacuum, which would imply a redefinition of the cosmological constant, without changing the overall picture. 
It is only for large spatial curvature fluctuations $v^2\lesssim ~^{(3)}R$ that our model would break down: the vacua of the bare potential would not be degenerate anymore, such that we would have to take into account the formation of bubbles of true vacuum inside the false vacuum.

\section{Background II: Homogeneous anisotropic universe description}
\label{Sec:bg_aniso_cosmo}

We review here features of the homogeneous but anisotropic universe relevant to our study, and in particular discuss how the anisotropy can be treated as an additional matter component in the Friedmann equations, 
assuming an appropriate equation of state. Starting with the anisotropic Bianchi-I metric
\be \label{eq:metric-BI}
\text{d}s^2=-\text{d}t^2+a^2_1\, \text{d}x^2+a^2_2\, \text{d}y^2+a^2_3\, \text{d}z^2~,
\ee
the Friedmann equations read ($i,j=1,2,3$)
\bea
H_1 H_2+H_1H_3+H_2H_3&=&+\kappa \rho~, \label{eq00}\\
(\mbox{for}~i\ne j)~~~~H_iH_j+\frac{\ddot a_i}{a_i}+\frac{\ddot a_j}{a_j}&=&-\kappa p ~ ,\label{eqii} 
\eea
where $H_i=\dot a_i/a_i$ and as above $\kappa=8\pi G$. Following \cite{Barrow:1995fn,Ellis:1998ct}
we note that eq.(\ref{eq00}) can be written
\be\label{eq00bis}
H^2=\frac{\kappa}{3}\rho+\sigma^2~,
\ee
where the averaged Hubble rate $H$ and the anisotropy $\sigma^2$ are defined as 
\bea
H&:=&\frac{1}{3}(H_1+H_2+H_3)~ ,\\
\sigma^2&:=&\frac{1}{18}\Big[(H_1-H_2)^2+(H_2-H_3)^2+(H_1-H_3)^2\Big]~.\nonumber
\eea
Equation (\ref{eq00bis}) shows that $3\kappa^{-1}\sigma^2$ can be interpreted as an energy density arising from anisotropy.
Similarly, the trace of Friedmann equations can be written
\be \label{eq:trace1}
\dot H+H^2=-\frac{\kappa}{6}(\rho+3 p)-2\sigma^2~,
\ee
such that $3\kappa^{-1}\sigma^2$ can also be interpreted as a pressure arising from anisotropy. Anisotropy therefore plays a role similar
to a homogeneous perfect fluid with equation of state $w=p/\rho=1$, and we expect that the corresponding energy density scales as $a^{-3(1+w)}=a^{-6}$, 
where $a=(a_1a_2a_3)^{1/3}$ is the average scale factor. This is consistent with Eqs. \eqref{eqii} from which one can show that
\be
\dot H_i-\dot H_j=-3H( H_i- H_j)~,
\ee
which implies $( H_i- H_j) \propto a^{-3}$ and thus $\sigma^2 \propto a^{-6}$.

Finally we note that, if a bounce occurs, then at this bounce $H=0$ and $\dot H>0$, such that the matter contribution should satisfy at the bounce
\bea \label{eq:NECconditions}
\rho&=&-\frac{3}{\kappa}\sigma^2\le0\\
\rho+3p&=&-\frac{6}{\kappa}\left(\dot H+2\sigma^2\right)\le0~,\nonumber
\eea
as in the isotropic case.

\section{Anisotropic quantum fluid description}
\label{Sec:aniso_qft}

As discussed above, in the isotropic case ($a_1=a_2=a_3\equiv a$) and from the effective potential (\ref{convexU}), the action for the true ground state $\phi=0$ is 
\be\label{Sisotropic}
S_{eff}^{isotropic}=\int \text{d}^4x\sqrt{-g}~\kappa^{-1}\Lambda\left(1-r~\frac{e^{-\alpha_{iso}^3}}{\alpha_{iso}^{3/2}}\right)~.
\ee
In the anisotropic case \eqref{eq:metric-BI}, the only change to the instanton action $\Sigma$ is via the volume $a^3V_{cell}\to a_1a_2a_3V_{cell}$,
such that the action (\ref{Sisotropic}) must be modified as
\be\label{Seff}
S_{eff}=\int \text{d}^4x\sqrt{-g}~\kappa^{-1}\Lambda\left(1-r~\frac{e^{-\alpha_1 \alpha_2 \alpha_3}}{\sqrt{\alpha_1 \alpha_2 \alpha_3}}\right)~,
\ee
where $\alpha_i\equiv a_i \Sigma^{1/3}$. The stress-energy tensor can be decomposed as   
\be 
T_{\mu \nu}\equiv\frac{2}{\sqrt{-g}}\frac{\delta S_{eff}}{\delta g^{\mu\nu}}=\textrm{diag}(\rho,a_1^2\,p,a_2^2\,p,a_3^2\,p)~ ,
\ee
and leads to the dimensionless energy density and pressure
\bea\label{rhotildeptilde}
\tilde\rho&\equiv&\frac{\kappa\rho}{\Lambda}=1-r~\frac{e^{-\alpha^3}}{\alpha^{3/2}}~,\\
\tilde p&\equiv&\frac{\kappa p}{\Lambda}=-1+r~\left(\frac{1}{2\alpha^{3/2}}-\alpha^{3/2}\right)e^{-\alpha^3}~, \nonumber
\eea
where the average scale factor $\alpha$ is defined by 
\be
\alpha^3\equiv a_1 a_2 a_3 \Sigma~.
\ee
In the previous expressions and from its definition in Eq.(\ref{rdef}), $r$ describes the quantum field -- it is completely determined once we specify its mass, vacuum energy and self interaction strength (via the parameters $v$, $\Lambda$ and $\lambda$). The pressure and energy density are therefore determined by the combination of field parameters $r$ and the average size of the universe $a\, V_{cell}^{1/3}$.

The Friedmann equations read, in terms of the rescaled quantities, with the rescaled time $\tau\equiv t~\sqrt{\Lambda/3}$,
\bea\label{rescaledFriedmann}
\frac{\mathcal{H}_1 \mathcal{H}_2}{3} +\frac{\mathcal{H}_1\mathcal{H}_3}{3}+\frac{\mathcal{H}_2\mathcal{H}_3}{3}&=&+\tilde \rho~, \label{eq0}\\
(\mbox{for}~i\ne j)~~~~\frac{\mathcal{H}_i\mathcal{H}_j}{3}+\frac{\alpha_i''}{3 \alpha_i}+\frac{\alpha_j''}{3\alpha_j}&=&-\tilde p ~ ,\label{eqjj} 
\eea
where a prime refers to the derivative with respect to $\tau$. 
We can then define the rescaled average Hubble rate $\mathcal{H}$ and anisotropy $\tilde\sigma^2$
\bea \label{eq:def:H:and:sigma}
\mathcal{H}&:=&\frac{1}{3}(\mathcal{H}_1+\mathcal{H}_2+\mathcal{H}_3)~ ,\\
\tilde\sigma^2&:=&\frac{1}{18}\Big[(\mathcal{H}_1-\mathcal{H}_2)^2+(\mathcal{H}_2-\mathcal{H}_3)^2+(\mathcal{H}_1-\mathcal{H}_3)^2\Big]~,\nonumber 
\eea
so that eq.(\ref{eq0}) can be simply written as
\be\label{Hsigmarho}
\mathcal{H}^2=\tilde\rho+\tilde \sigma^2~.
\ee
We can see from eq. \eqref{eq:NECconditions} that $\tilde\rho<0$ and $\tilde\rho+3\tilde p<0$,
in order to balance out the anisotropy contribution in the vicinity of the average bounce, defined by $\mathcal{H}=0$ and $\mathcal{H}'>0$.
In what follows, we will study under which conditions the cosmological bounce can be induced.

\section{Critical solutions}\label{Sec:analytic}

As discussed in the previous section, in a universe with significant anisotropy, an additional contribution to the energy density and the pressure of the spacetime exists. Starting from some initial condition, several scenarios are possible given the different scalings in $\alpha$. 
The NEC violation from tunnelling does not necessarily win over the anisotropy during the collapse (even where it is initially larger) a bounce requires not only that both contributions cancel each other 
such that $\mathcal{H}=0$, but also that at this point of equality, the pressure satisfies the necessary condition for the universe to bounce (i.e., $\mathcal{H}'>0$). For this latter condition to be true, the size of the universe at this point must be sufficiently small (relative to the field parameters) for finite volume effects to be significant, but not too small to avoid a collapse. In this section we derive the specific requirements, and comment on the resulting phenomenology.

\subsection{Critical point}

The critical solution of the Friedmann equations for which a bounce occurs $(\rho_c,p_c,\alpha_c,\tilde\sigma_c^2)$ can be found by imposing the condition $\mathcal{H}=\mathcal{H}'=0$. This critical point is unstable:
a value of $\alpha$ that is slightly larger than $\alpha_c$ leads to a bounce (the NEC violation $\propto\alpha^{-3/2}$ dominates) and a value which is slightly smaller leads to a collapse (the anisotropy $\propto\alpha^{-6}$ dominates). 

From these conditions, the energy density and the pressure at the critical point satisfy
\be
\tilde p_c= \tilde \rho_c=-\tilde \sigma^2_c~.
\ee
Also, from Eqs. (\ref{rhotildeptilde}), we find that 
the averaged scale factor $\alpha_c$ is given by the implicit algebraic equation
\be \label{eq:algebraic:alpha:limit}
4\alpha_c^{3/2}+r e^{-\alpha_c^3}(-3+2 \alpha^3_c)=0~,
\ee
and the anisotropy can be expressed as
\be\label{sigmal}
\tilde \sigma^2_c=\frac{1+2 \alpha^3_c}{3-2\alpha^3_c}~.
\ee
We can see from Eq. (\ref{eq:algebraic:alpha:limit}) that necessarily $\alpha_c^3<3/2$, and
one can identify the two regimes
\bea\label{limitsalphal}
\alpha_c&\to&(3/2)^{1/3}~~~~\mbox{for}~~r\gg1\\
\alpha_c&\sim&\left(\frac{3r}{4}\right)^{2/3}~~~~\mbox{for}~~r\ll1~.\nonumber
\eea

One can understand the role of $\alpha_c$ from the point of view of the pressure. Assume that there is a time $\tau_1$ where $\mathcal{H}(\tau_1)=0$:
\begin{itemize}
    \item  A bounce requires the condition $\mathcal{H}'(\tau_1)>0$, and thus $|\tilde p(\tau_1)|>|\tilde\rho(\tau_1)|$, such that
\be
4\alpha^{3/2}(\tau_1)+re^{-\alpha^3(\tau_1)}[-3+2\alpha^3(\tau_1)]>0~,
\ee
which leads to $\alpha(\tau_1)>\alpha_c$;\\
    \item  A collapse follows in the situation where $\mathcal{H}'(\tau_1)<0$, and thus $|\tilde p(\tau_1)|<|\tilde\rho(\tau_1)|$, such that
\be
4\alpha^{3/2}(\tau_1)+re^{-\alpha^3(\tau_1)}[-3+2\alpha^3(\tau_1)]<0~, 
\ee 
which leads to $\alpha(\tau_1)<\alpha_c$.
\end{itemize}

\subsection{Comparison with the isotropic case}

\begin{figure}[h!]
    \centering
    \includegraphics[width=8cm]{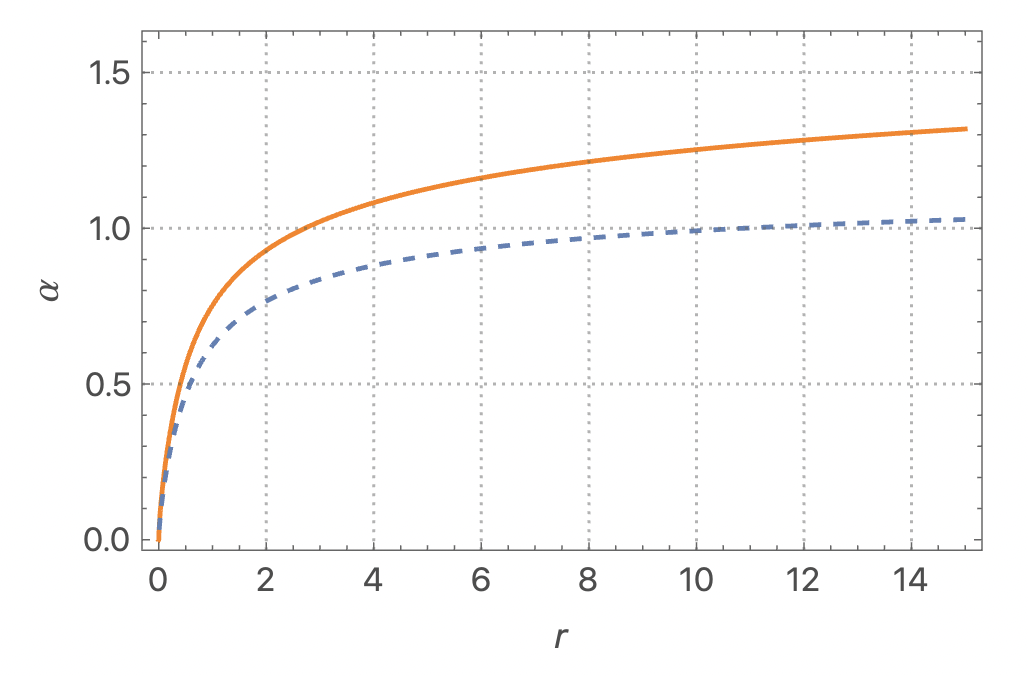}
    \caption{\justifying{\small A plot of the rescaled average scale factor at the bounce $\alpha = a \Sigma^{1/3}$ (which is related to the size of the closed universe) versus $r$ (which is determined by the properties of the field). We plot the maximum $\alpha_{iso}$ (orange line) and minimum $\alpha_c$ (dashed blue line) values, as a function of $r$. Although $\alpha_c$ asymptotically tends to $(3/2)^{1/3}$,
    $\alpha_{iso}$ is not bounded and goes to infinity when $r\to\infty$. }}
    \label{fig:alphai-alphaf}
\end{figure}

One can also infer a maximum value for the rescaled scale factor $\alpha_{iso}$ at the bounce, which happens when the anisotropy (and any other components if there are) are negligible and the quantum field completely dominates.
We then have from Eq. (\ref{rhotildeptilde})
\be
\alpha_{iso}^{3/2}=r~e^{-\alpha_{iso}^3}~,
\ee
which leads to the two regimes
\bea
\alpha_{iso}&\sim&(\ln r)^{1/3}~~~~\mbox{for}~~~~r\gg1\\
\alpha_{iso}&\sim&r^{2/3}~~~~\mbox{for}~~~~r\ll1~,
\eea
and we note that $\alpha_{iso}$ is not bounded when $r\to\infty$.
We sketch $\alpha_c$ and $\alpha_{iso}$ on Fig.\ref{fig:alphai-alphaf}, where the region between the two curves represents the possible range of values of the rescaled scale factor at which a bounce can occur for a particular quantum field (as parametrised by $r$).

\subsection{Size of the Universe at the bounce}

\begin{figure}[h!]
    \centering
   \includegraphics[width=0.95\columnwidth]{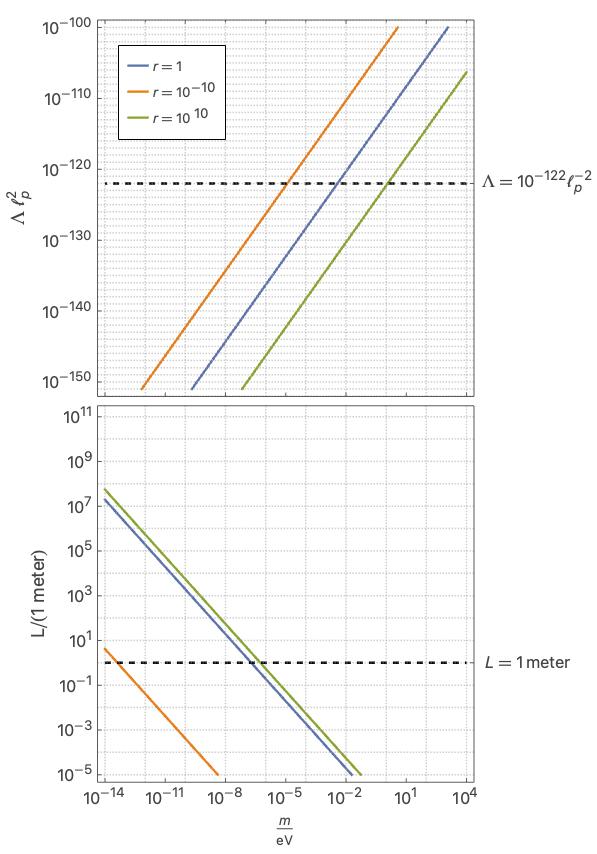}
 
    \caption{\justifying{\small{ These plots illustrate consistent values of the field parameters and the size $L_b$ of the closed universe at the bounce for different values of the field parameter $r$. The upper plot shows the relation between $\Lambda$ and $m$. 
    The lower plot shows the relation between the maximum value for $L_b$ and $m$ 
    (the minimum of $L_b$ has very similar values and is not represented for the sake of clarity - thus the range of possible values between the limits in Fig. \ref{fig:alphai-alphaf} all lie close to the lines in this plot).    
    We note that the function $L_b(m)$ is almost independent of $r$ for $r\gg1$, whereas it changes significantly for $r\ll1$.
    }}}
    \label{fig:pheno}
\end{figure}

From the previous results one can put bounds on the typical physical size of the universe at the bounce, for a given field. 
The instanton action is of the order \cite{Alexandre:2023iig}
\be 
\Sigma\sim\frac{m^3}{\lambda}V_{cell}~,
\ee
where $m=v\sqrt{\lambda/3}$.
The physical length $L$ is given by $L=a(t)V_{cell}^{1/3}$ and its value at the bounce then satisfies
\be
\lambda^{1/3}~\frac{\alpha_c}{m}~\lesssim~ L_b~\lesssim~\lambda^{1/3}~\frac{\alpha_{iso}}{m}~.
\ee

To get a feel for the phenomenological consequences of the model, we can relate the field quantities $m$ and $\Lambda$ with the size of the closed universe at the bounce $L_b$, parameterised by the ratio $r$. 
For simplicity we assume here that $\lambda^{1/3}$ is of order 1
\footnote{This assumption is not necessarily justified for an axion-like particle, with a potential of the form $M^4\cos(\phi/f)$. Indeed, a small mass $\propto M^2/f$ compared to 1eV \cite{Marsh:2015xka} implies an extremely small self-coupling constant $\propto M^4/f^4$. However, there are axion models not requiring such a small self coupling, as for example in the string-inspired model presented in \cite{Mavromatos:2023bdx}. Our results can easily be adapted to other values for $\lambda$ depending on the model.}.

The current size of the visible universe is $\sim10^{26}$ meters, and we do not see any evidence of periodicity in it \cite{baumann_2022}. Any bounce must have happened before the electroweak phase transition, at which point the size of our observable universe was about $10^{11}$ meters. This therefore imposes a minimum on the size of the closed universe at the bounce - any smaller and we would see evidence for periodicity now. However, a bounce could also have occurred much earlier than this and so the universe could have been smaller. If instead we take the bounce to occur at the era of grand unification, the size of the closed universe would be of order $1$ meter or larger.

As can be seen from Fig.\ref{fig:pheno}, for $r=1$, if the bounce occurred when $L \sim 1$ m, the scalar field would need a bare mass of $\sim 10^{-7}$ eV and a vacuum energy $\Lambda$  of order $10^{-140} ~ \ell_p^{-2}$, therefore much smaller than the current cosmological constant. The plot illustrates how the values change for different values of $r$, but in general one requires a larger mass to be consistent with a smaller $L$, and small values for the vacuum energy are required.

\section{Numerical solutions}\label{Sec:numeric}

In this section we numerically integrate the Friedmann equations (\ref{rescaledFriedmann}), both in the bouncing case (where NEC violation dominates) and in the collapsing case (where the anisotropy dominates), to illustrate the possible outcomes. 

We choose the initial anisotropy $\tilde\sigma^2(\tau_0)$ as one of the parameters. For simplicity, we will consider $\alpha_2(\tau_0)=\alpha(\tau_0)$ and 
 $\mathcal H_2(\tau_0)=\mathcal H(\tau_0)$. Hence from \eqref{eq:def:H:and:sigma} we find for all times
\bea \label{eq:H1:sigma}
\mathcal{H}_1&=&\mathcal{H}\pm \sqrt{3\tilde \sigma^2}~,\\
\label{eq:H2:sigma}\mathcal{H}_2&=&\mathcal{H}~,\\
\mathcal{H}_3&=&\mathcal{H}\mp \sqrt{3\tilde \sigma^2}~ . \label{eq:H3:sigma}
\eea
The initial value of $\mathcal{H}$ can then be determined by equations \eqref{Hsigmarho} and (\ref{rhotildeptilde}), namely
\be\label{eq:constraint:initial}
\mathcal{H}^2(\tau_0)=\tilde\sigma^2(\tau_0)+1-r \frac{e^{-\alpha^3(\tau_0)}}{\alpha^{3/2}(\tau_0)}~.
\ee
We take the negative root $\mathcal{H}(\tau_0)<0$ since we want to start from a contracting phase. 
As a consequence, the initial Hubble rates are entirely determined by $\tilde \sigma^2(\tau_0)$, and for the numerical analysis the quantities we fix are $\tilde \sigma^2(\tau_0)$, $\alpha(\tau_0)$ and $r$.

We take $r\leq e$, so that we are before the point of equality in the energy densities in the anisotropy and the quantum field. 
For these values of $r$, we choose the initial average scale factor such that $\alpha^3(\tau_0)=\alpha_1(\tau_0)\alpha_2(\tau_0)\alpha_3(\tau_0)=1$. 
This allows the bounce to happen soon after the initial time, compared to the typical time scale of the whole process. A larger initial 
scale factor would shift the time when the bounce occurs.

\begin{figure}[h] 
 \includegraphics[width=7.5cm]{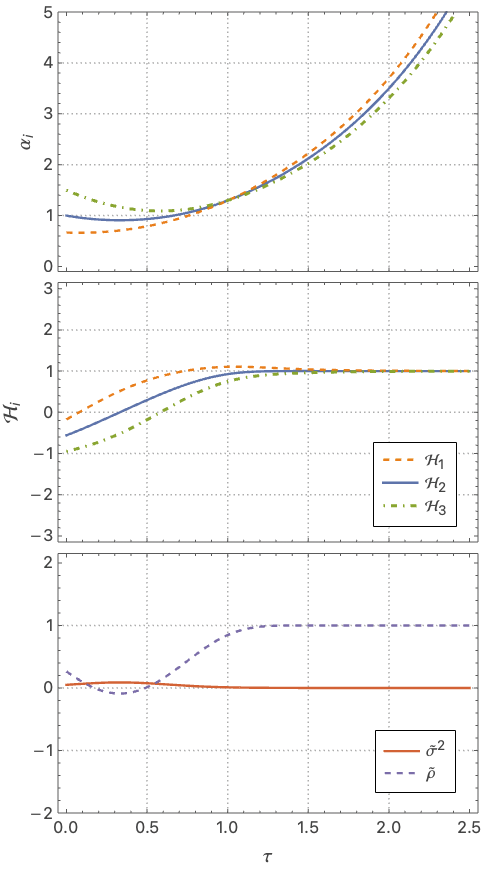}
        \caption{\justifying{\small{{\bf Bouncing case:} In the two upper panels we show the time evolution of the rescaled scale factors $\alpha_i$ 
        and the rescaled Hubble rates $\mathcal{H}_i$ with the initial conditions given in the main text.  The lower panel shows
        the time evolution of the anisotropy $\tilde \sigma^2$  and energy density $\tilde \rho$.}}}
        \label{fig:A}
\end{figure}

Fig.\ref{fig:A} shows an example of a bouncing solution. We include the time evolution of the scale factors $\alpha_i$, 
and Hubble rates $\mathcal{H}_i$, the anisotropy $\tilde \sigma^2$, and the energy density $\tilde \rho$.  
We choose $r=2$, and initial conditions at $\tau_0=0$ $\alpha_i(\tau_0)=\{2/3,1,3/2\}$, and $\tilde\sigma^2(\tau_0)=0.05$. 

\begin{figure}[h] 
 \includegraphics[width=7.5cm]{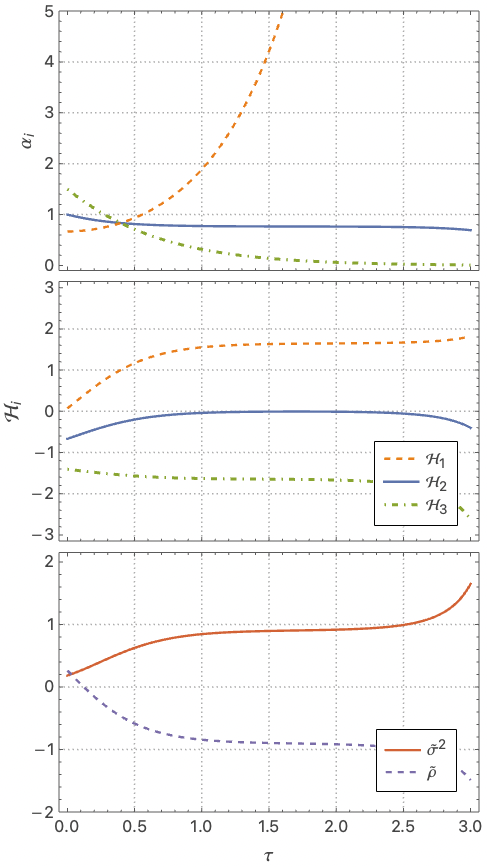}
        \caption{\justifying{\small{{\bf Near critical no-bounce case:} In the two upper panels we show the time evolution of the rescaled scale 
        factors $\alpha_i$ and the rescaled Hubble rates $\mathcal{H}_i$ with the initial conditions given in the main text.  
        The lower panel shows the time evolution of the anisotropy $\tilde \sigma^2$ 
        and energy density $\tilde \rho$. The initial conditions are close to the critical point and for a time the value of the rescaled average Hubble rate $\mathcal{H}$ remains near zero. However, eventually the average scale factor grows, leading to another collapsing phase in which the anisotropy diverges.}}}
        \label{fig:C}
\end{figure}

Fig.\ref{fig:C} shows an example of a case where the bounce is not reached, due to the anisotropy dominating, but that is close to the critical case. 
The initial conditions are $r=2$,  $\alpha_i(\tau_0)=\{2/3,1,3/2\}$ and $\tilde\sigma^2(\tau_0)=0.18247$.

\section{Discussion}
\label{Sec:discuss}

In this work we have studied the possibility of a cosmic bounce occurring in a universe in which there is a significant (but not dominant) anisotropy, in addition to the presence of a scalar field which is subject to finite volume effects in a closed universe. We have shown that criteria for the success of the bounce can be stated in terms of the size of the closed universe at the point at which the net energy density is zero, and have studied the properties of the quantum field (mass and vacuum energy) that permit a bounce of a size consistent with our own cosmological history. 

At the point of zero net energy density, the size of the universe must be roughly comparable to the Compton wavelength of the field for its pressure to be sufficient to turn around the collapse, so values smaller than $\sim 10^{-5}$ eV are needed for the mass. We also find that the vacuum energy of the field must be extremely small, even in comparison to the current day cosmological constant. 
After the bounce, the universe transitions to expansion and the contribution of the quantum field to the energy density reduces to its vacuum energy, with finite volume effects completely suppressed. The fact that this vacuum energy is smaller than the current value is therefore consistent with what we observe (it could be a small contribution to its value), but the smallness of the value seems to require some further explanation -- although this is of course true of the cosmological constant itself.

Examples of further consequences of tunnelling effects in finite volume are as follows:
One could involve an out-of-equilibrium QFT description of tunnelling in the background of a time-varying metric, which would allow a more accurate study away from the bounce. Then, the inclusion of the Casimir effect due to the finite volume $V_{cell}$ could give rise to new effects, with possible cosmological relevance. Also, the effect of spatial curvature should be included, in the situation where the space fundamental cell is a 3-sphere instead of a 3-torus. These studies are left for future works.

\acknowledgments

We thank Tim Clifton, Malcolm Fairbairn and David J. Marsh for helpful conversations. 
This work is supported by the Leverhulme Trust (grant No. RPG-2021-299). 
J.A. is also supported by the Science and Technology Facilities Council (grant No.  STFC-ST/T000759/1). 
K.C. is supported by an STFC Ernest Rutherford fellowship, project reference No. ST/V003240/1. 
For the purpose of Open Access, the authors have applied a CC BY public copyright licence to any Author Accepted Manuscript version arising from this submission.

\bibliography{biblio}

\end{document}